# Parity-time symmetry from stacking purely dielectric and magnetic slabs

James Gear[1], Fu Liu[1], S. T. Chu[2], Stefan Rotter[3] and Jensen Li[1]*

[1] *School of Physics and Astronomy, University of Birmingham, Birmingham B15 2TT, UK*
[2] *Department of Physics and Materials Science, City University of Hong Kong, Tat Chee Avenue, Kowloon Tong, Hong Kong*
[3] *Institute for Theoretical Physics, Vienna University of Technology, A-1040 Vienna, Austria, EU*

We show that Parity-time symmetry in matching electric permittivity to magnetic permeability can be established by considering an effective Parity operator involving both mirror symmetry and coupling between electric and magnetic fields. This approach extends the discussion of Parity-time symmetry to the situation with more than one material potential. We show that the band structure of a one-dimensional photonic crystal with alternating purely dielectric and purely magnetic slabs can undergo a phase transition between propagation modes and evanescent modes when the balanced gain/loss parameter is varied. The cross-matching between different material potentials also allows exceptional points of the constitutive matrix to appear in the long wavelength limit where they can be used to construct ultrathin metamaterials with unidirectional reflection.

* j.li@bham.ac.uk

Parity-time (PT) symmetric Hamiltonians are proposed as an extended class of Hamiltonians in quantum mechanics, in addition to the conventional Hermitian ones [1, 2]. On the one hand, these PT-symmetric Hamiltonians can give rise to real eigenvalues, similar to the Hermitian ones. On the other hand, when the system is subject to changes through certain external parameters, PT-symmetric Hamiltonians can transit from the PT-symmetric phase to the PT-broken phase, which gives rise to pairs of complex conjugate eigenvalues beyond a certain threshold. Such a PT-phase transition and the associated threshold, called an exceptional point, provide a much wider platform for exploring extraordinary quantum mechanical phenomena. However, these PT-symmetric Hamiltonians are difficult to implement in a quantum mechanical system, requiring $V(x) = V^*(-x)$ in balancing gain and loss. In fact, other analog systems are being used to explore the interesting physics of PT-symmetric systems and of exceptional points in general, including mechanical and optical systems [3-7]. Currently, there is a tremendous interest in investigating PT-symmetry in optics since it can give rise a wide range of counterintuitive wave phenomena. It should be noted that the PT-symmetric potentials studied generally fall into two categories: those with symmetry perpendicular to the propagation direction, where behavior is derived from the paraxial wave approximation, and those with symmetry parallel to the propagation direction, where wave continuity or transfer matrix methods are used for analysis. For those with perpendicular symmetry: novel beam refraction [8], power oscillation [9], loss-induced transparency [10], nonreciprocal Bloch oscillations [11], non-reciprocal or phase-transition in scattering [12, 13] have been observed, for instance. For those with parallel symmetry: laser absorber modes [14,15], unidirectional invisibility [16-18] have been observed. In addition, these investigations are also extended to various extraordinary lasing and nonlinear effects with resonating structures [19-25]. In all of these cases, an analog condition of PT-symmetry for the optical refractive index profile is used: $n(x) = n^*(-x)$ where $x$ is taken as the direction either along or perpendicular to the propagation direction.

In parallel to the above development, metamaterials, consisting of an array of artificial atoms on the subwavelength scale, are being used as a flexible platform to obtain intriguing optical properties frees from the limited palette of constitutive relationships found in natural materials. Metamaterials with negative refractive indices, bianisotropic tensors and chiral parameters are now routinely fabricated [26-28]. Similarly, we also expect metamaterials to be flexible enough to engineer the Hamiltonian of an optical system and hence to investigate PT-symmetry [29-34]. A natural extension to discuss PT-symmetry in the framework of metamaterials is to use the spatial inversion or mirror operator as the parity operator for matching both the permittivity profile or the permeability profile [30, 31]. In fact, when there is more than one material potential, additional degrees of freedom exist in defining the PT symmetry. In this work, we will explore the possibility of "cross"-matching different material potentials (matching $\epsilon(x)$ to $\mu^*(-x)$) in defining the PT symmetry, giving us additional flexibility in designing PT-symmetric components.

**Parity-time symmetry operator coupling electric and magnetic fields**

We consider a one dimensional wave problem in $z$-direction for which both the materials and the fields are invariant in the transverse direction (taken as $x$- and $y$-directions). In such a case, the parity operator can be taken as $P = M_z$, i.e. the mirror operation along the $z$-direction, while the $T$ operator is taken as the time-reversal operation. To achieve PT-symmetry, one conventionally considers just the profile of the one dimensional permittivity $\epsilon(z)$, which then has to satisfy the condition $\epsilon(z) = \epsilon^*(-z)$. If, however, one also incorporates a magnetic permeability profile $\mu(z)$ on top of the permittivity profile, one expects an additional but similar requirement on the magnetic permeability: $\mu(z) = \mu^*(-z)$ [30]. It can also be written in terms of index $n(z) \triangleq \sqrt{\epsilon(z)}\sqrt{\mu(z)} = n^*(-z)$ and impedance $\eta(z) \triangleq \sqrt{\mu(z)}/\sqrt{\epsilon(z)} = \eta^*(-z)$. In both cases these PT-symmetric systems have the appealing property that they behave similarly to a lossless system in one phase while displaying entirely new features in the PT-broken phase. This PT-phase transition with the associated exceptional point separating the two phases, are emerging as a unique way to achieve tunable components with extreme sensitivity and unconventional behavior [12,22-25].

An important point to emphasize here is that the above PT-symmetry condition on the material profiles is derived with respect to a particular choice of the $P$ operator. Due to the recent capability in controlling magnetic permeability profile through metamaterials, we seek here for alternative parity operators by exploring the dual operation between the electric ($\vec{E}$) and the magnetic field ($\vec{H}$). In this work, we focus on the following definition of $P$ given by

$$P\begin{pmatrix}\vec{E}\\\vec{H}\end{pmatrix} = \begin{pmatrix} 0 & i\hat{z} \times I \\ -i\hat{z} \times I & 0 \end{pmatrix} M_z \begin{pmatrix}\vec{E}\\\vec{H}\end{pmatrix}, \tag{1}$$

where $M_z$ is the mirror operator in the $z$-direction and $I$ is the identity operation. In this construction we combine the ordinary parity operator with an additional operator that cross-couples $\vec{E}$ and $\vec{H}$. This combined operator is designed to possess the various properties of an ordinary parity operator as satisfying $P^2 = I$ and being unitary, hermitian and being a symmetry operator of the vacuum Maxwell's equations. In this sense, it can be regarded as an effective parity operator, which, for simplicity, we will just call the parity operator hereafter. The time-reversal operator, on the other hand, is written as

$$T \begin{pmatrix} \vec{E} \\ \vec{H} \end{pmatrix} = \begin{pmatrix} I & 0 \\ 0 & -I \end{pmatrix} \begin{pmatrix} \vec{E}^* \\ \vec{H}^* \end{pmatrix}. \tag{2}$$

By demanding an electromagnetic field and its PT-transformed version are both solutions of the same set of wave equations (within Heaviside-Lorentz units):

$$\partial_z \begin{pmatrix} H_y \\ E_x \end{pmatrix} = ik_0 \begin{pmatrix} \epsilon(z) & 0 \\ 0 & \mu(z) \end{pmatrix} \begin{pmatrix} E_x \\ H_y \end{pmatrix}, \tag{3}$$

with $k_0$ being the wave number in free space, we obtain the PT-symmetry condition on the current case of isotropic $\epsilon(z)$ and $\mu(z)$ as

$$\epsilon(z) = \mu^*(-z). \tag{4}$$

This condition will be employed to construct PT-symmetric material profiles in this work. Again, the same condition can also be written in terms of $n$ and $\eta$ by $n(z) = n^*(-z)$ and $\eta(z) = 1/\eta^*(-z)$. While the condition on the index is expected, the reciprocal condition on the impedance is in contrast to the ordinary one: $\eta(z) = \eta^*(-z)$.

To be more specific, we consider the material profile in Eq. (4) for a system, which is periodic with a lattice constant $a$. The fields on the two sides of a single unit cell, $E_{x1}, H_{y1}$ on the left and $E_{x2}, H_{y2}$ on the right hand side, can then be solved from Eq. (3) and are related to each other using

$$\begin{pmatrix} H_{y2} - H_{y1} \\ E_{x2} - E_{x1} \end{pmatrix} = \frac{ik_0 a}{2} C^{-1} \begin{pmatrix} E_{x2} + E_{x1} \\ H_{y2} + H_{y1} \end{pmatrix}. \tag{5}$$

This relation can be regarded as a discretized wave equation with $C^{-1}$ being defined as the constitutive matrix of the corresponding photonic crystal. Equivalently, we can also write the constitutive matrix using the scattering matrix $S$ of a single unit cell by

$$C^{-1} = \frac{2}{ik_0 a} B \frac{S - I}{S + I} B^{-1}, \text{with } B = \begin{pmatrix} 1 & 1 \\ 1 & -1 \end{pmatrix}. \tag{6}$$

Here, we have employed the convention of S-matrix defined by

$$S = \begin{pmatrix} t & r_b \\ r_f & t \end{pmatrix}. \tag{7}$$

where $t$ is the transmission amplitude and $r_f(r_b)$ is the reflection amplitude in the forward [left to right] (backward [right to left]) direction. Such a $C^{-1}$ matrix inherits the PT-symmetry from the microstructural profile. By considering the PT-operation on the black-box description in Eq. (5) with reciprocity, we obtain the form of a PT-symmetric $C^{-1}$ as

$$C^{-1} = \begin{pmatrix} \epsilon_{\text{eff}} & i\kappa_{\text{eff}} \\ -i\kappa_{\text{eff}} & \mu_{\text{eff}} \end{pmatrix}. \tag{8}$$

with

$$\epsilon_{\text{eff}} = \mu_{\text{eff}}^* \text{ and } \kappa_{\text{eff}} = \kappa_{\text{eff}}^*. \tag{9}$$

Here, $\epsilon_{\text{eff}}(\mu_{\text{eff}})$ is the effective permittivity (permeability) while the bianisotropy term $\kappa_{\text{eff}}$ arises from the effect of a finite frequency. Actually, this $C^{-1}$ is a well-defined system response matrix at an arbitrary frequency. This definition converges to the usual effective constitutive matrix only when the effective medium approximation is valid in the long wavelength limit [35].

For an ordinary PT-symmetric profile, i.e. $\epsilon(z) = \epsilon^*(-z)$, the gain and loss average out to zero in the effective medium regime so that the resultant effective medium only possesses trivial PT-symmetry (Hermitian $C^{-1}$ with purely real diagonal terms). In contrast, our PT-symmetric effective medium still possesses non-trivial PT-symmetry (non-Hermitian $C^{-1}$ with imaginary diagonal terms). As we shall see, exceptional points and the PT-phase transition are therefore possible even in the long wavelength limit. We note that the medium with isotropic $\epsilon$ and $\mu$ profiles in consideration guarantees decoupling into transverse electric and transverse magnetic polarizations. Therefore, Eqs. (3) to (8) are only specified for one polarization. All the results in this work can be extended to anisotropic media (e.g. with anisotropy in the x- and y-directions) in a straightforward manner where the material parameters are defined with respect to a particular polarization.

**Band structure and PT-phase transition**

To demonstrate the physical properties of the proposed PT-symmetric medium with dual symmetry between electric and magnetic fields, we start by investigating the band structure of a one dimensional photonic crystal with PT-symmetry and lattice constant $a$. To facilitate the discussion, we introduce the propagation matrix $K$ through the following definition:

$$\begin{pmatrix} E_{x2} - E_{x1} \\ H_{y2} - H_{y1} \end{pmatrix} = \frac{ia}{2} K \begin{pmatrix} E_{x2} + E_{x1} \\ H_{y2} + H_{y1} \end{pmatrix}. \tag{10}$$

where $K$ is related to $C^{-1}$ by

$$K = k_0 \begin{pmatrix} 0 & 1 \\ 1 & 0 \end{pmatrix} C^{-1}. \tag{11}$$

Then, the dispersion of the photonic crystal can be written as an eigenvalue problem of $K$ as

$$K \begin{pmatrix} E_x \\ H_y \end{pmatrix} = \frac{2}{a} \tan \frac{ka}{2} \begin{pmatrix} E_x \\ H_y \end{pmatrix}, \tag{12}$$

where $\exp(ika)$ is the Bloch phase factor of the eigen-mode. This eigenvalue problem (which can be equivalently formulated using the transfer matrix) defines different PT-phases and the exceptional points which separate them. From Eq. (8) and Eq. (9) and the definition of $K$ matrix in Eq. (11), one can derive the dispersion equation for a PT-symmetric system becomes

$$\pm \frac{2}{k_0 a} \tan \frac{ka}{2} = \sqrt{|\epsilon_{\text{eff}}|^2 - \kappa_{\text{eff}}^2}. \tag{13}$$

Therefore, the Bloch $k$ can either come in pairs of two purely real numbers of opposite sign when $|\epsilon_{\text{eff}}|^2 > \kappa_{\text{eff}}^2$ or of two (conjugate) purely imaginary numbers when $|\epsilon_{\text{eff}}|^2 < \kappa_{\text{eff}}^2$. In this sense, the propagation matrix $K$ plays the similar role to the Hamiltonian whose eigenvalue problem, bandstructure here, is a valid concept that naturally extends from the lossless case (PT-symmetric phase being lossless-like with two real eigenvalues). The right hand side of Eq. (13) can be interpreted as an interaction parameter in band formation [36,37]. A real/zero/imaginary interaction parameter corresponds to propagation band/exceptional point/band gap in a band-structure.

As an example, we investigate in the following a PT-symmetric photonic crystal that is constructed by stacking two types of media with equal thickness. One of them is a purely dielectric slab with $\epsilon_1 = 2.25 + i\gamma$, $\mu_1 = 1$ and the other one is a purely magnetic slab with $\mu_2 = 2.25 - i\gamma$, $\epsilon_2 = 1$ (profiles shown in Fig. 1(a)). Such a photonic crystal satisfies the PT-symmetric condition listed in Eq. (4). In general, we can also choose slabs with both electric and magnetic responses but we focus on purely dielectric and purely magnetic slabs to demonstrate the ability to cross-match different types of potentials to establish PT symmetry. The band structure (frequency against real part of $k$) is plotted in Fig. 1(b) as $\gamma$ varies. When $\gamma = 0$ (the lossless case), the band structure (highlighted by black solid lines) is a conventional one with the propagating bands transversing from $k = 0$ to $k = \pi/a$ at the Brillouin zone edge where photonic band gaps open. The band structure for negative $k$ is the mirror image of the one for positive $k$. As $\gamma$ increases from zero, the propagating bands with two real $\cot(ka/2)$ values gradually close up to zero at the exceptional point. Beyond the exceptional point, two purely imaginary $\cot(ka/2)$ values emerge. The band structure undergoes a PT-phase transition beyond which the bands become evanescent with nonzero $Im(k)$ and $Re(k) = \pi/a$. This is the PT-broken phase, or the generalized band gap, represented by the deep blue region in Fig. 1(b) if we regard the top view of this diagram as a phase diagram. The $\gamma$ required to enter the broken phase, approximately scales with $1/\omega$, decreases as we go to a higher working frequency.

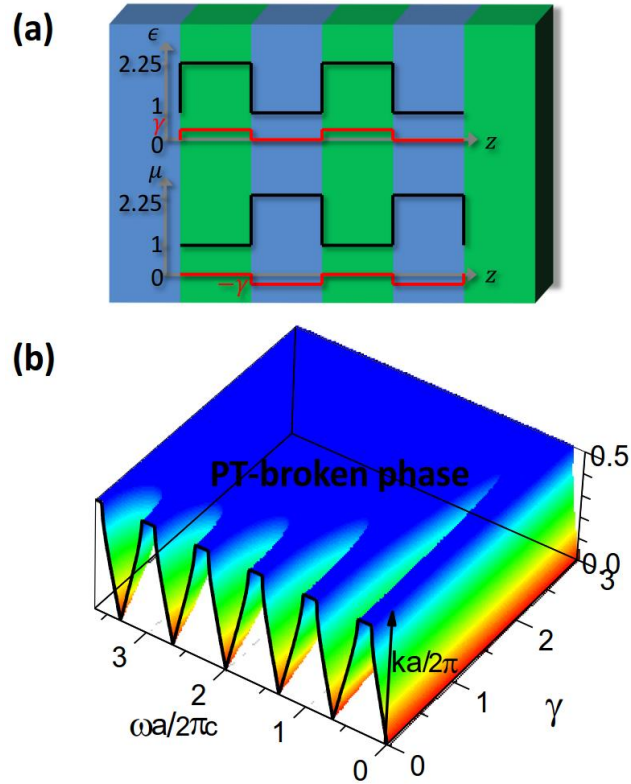

Fig. 1. (Color online) (a) The permittivity ($\epsilon$) and permeability ($\mu$) profiles of a one-dimensional photonic crystal stacking alternating purely dielectric slab of $\epsilon = 2.25 + i\gamma$ and purely magnetic slab $\mu = 2.25 - i\gamma$ with equal filling ratio and varying $\gamma$. (b) The corresponding band structure: Bloch $k$ against normalized frequency and $\gamma$. The top surface of the plot labels the PT-broken phase (generalized band gap).

The band structure and the PT-symmetric phases can also be revealed through the scattering behavior of a photonic crystal with a finite number of unit cells in thickness. Figure 2 plots the reflectance ($|r_f|^2$ in the forward $|r_b|^2$ in the backward direction) of the same photonic crystal for 4 unit cells. The reflection amplitude oscillates within the bands (PT-symmetric phase) with dips and peaks from Fabry-Perot resonance while it stays high in the gap region (PT-broken phase). This clearly identifies the two separate phases of PT-symmetry by the reflectance in both directions. By comparing the forward and backward reflectance, we find that the reflectance from the side with loss actually constitutes additional lines of dip (in red color in Fig. 2(a)). These dips, reaching value zero, display a unidirectional character in reflection, which will be investigated later in this work.

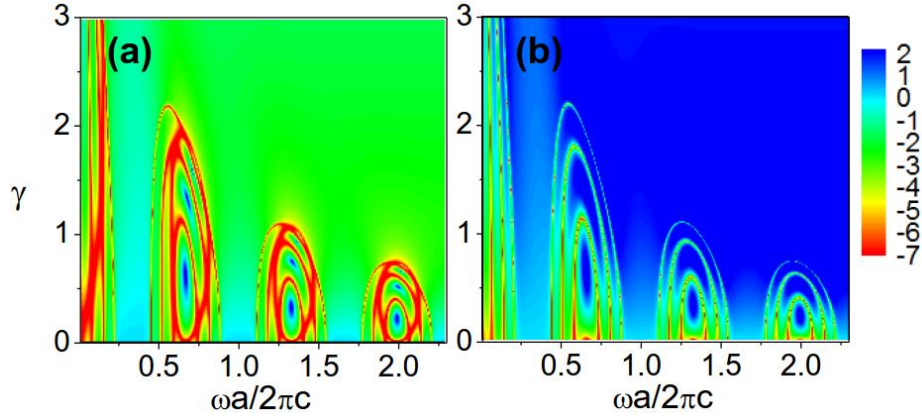

Fig. 2. (Color online) (a) Forward and (b) backward reflectance (in natural log scale) for the photonic crystal in Fig. 1 with a finite thickness of 4 unit cells.

We note that in the examples shown in this work, we have chosen the simple additive model to incorporate the balanced gain/loss parameter $\gamma$ ($i\gamma$ being added to or subtracted from $\epsilon$ or $\mu$) in the constitutive paramters $\epsilon_1$ and $\mu_2$. It is just a mathematical convenience for simplicity to demonstrate the various implications of the proposed PT-symmetry. The phase transition behavior depends on the actual microscopic model on how gain and loss are added. In such a case, we can seek particular frequencies to fulfill PT-symmetry while the various formulas in this work (e.g. the eigenvalue problem of the dispersion is formulated at a fixed frequency) are still valid.

**Unidirectional character in reflection in the effective medium limit**

The unidirectional character in the reflection amplitude for the demonstrated photonic crystal ($r_f = 0$ with nonzero $|r_b|$) occurs at the exceptional point (eigenvalue degeneracy) of the scattering matrix. This phenomenon turns out to be invariant with the number of unit cells (appearing at the same frequency and $\gamma$ values) so that we can simply employ the $S$ matrix of a single unit cell defined in Eq. (7) for studying it. By investigating the eigenvalue behavior of the $S$ matrix, one can identify the exceptional point where this unidirectional character occurs. As the $S$ matrix of a single unit cell is related to the constitutive matrix $C^{-1}$ through the bilinear transformation shown in Eq. (6), we can equivalently study the eigenvalue behavior of $C^{-1}$ of the photonic crystal. Again, from Eq. (8) and Eq. (9), one can immediately see the eigenvalues of a PT-symmetric $C^{-1}$ (denoted as $c_i^{-1}$ with $i = 1,2$) can either come in pairs of two purely real numbers or of two (conjugate) purely imaginary numbers. This eigenvalue behavior (similar to a PT-symmetric Hamiltonian in quantum mechanics) justifies our identification of $C^{-1}$ of the crucial quantity for discussing the unidirectional character in reflection. Generally speaking, different optical phenomena (with exceptional points or with PT-phase transition) link to different eigenvalue problems. The corresponding identification is, however, not unique. Here, we have linked the dispersion problem (solving $k$ with fixed $\omega$) to the eigenvalue problem of the propagation matrix and unidirectional reflection to the eigenvalue problem of the constitutive matrix.

Figure 3(a) shows a figure of merit for the degree of directivity $\chi$ ($\chi = 1$ for completely unidirectional and $\chi = 0$ for completely symmetric behavior) with the definition from the S-matrix of a single unit cell:

$$\chi = \left| \frac{|S_{12}|^2 - |S_{21}|^2}{|S_{12}|^2 + |S_{21}|^2} \right|. \tag{14}$$

In Fig. 3(a), the value of $\chi$ approaches one (at the two black solid lines) for a conventional PT-symmetric system by matching two dielectric slabs $\epsilon_1 = 2.25 + i\gamma$ and $\epsilon_2 = 2.25 - i\gamma$ within a unit cell. Figure 3(b) shows the corresponding results for our proposed PT-symmetry, with the second slab replaced by $\mu_2 = 2.25 - i\gamma$ with unit permittivity. Here, we are particularly interested in the unidirectional behavior at the long wavelength limit. As we have mentioned, the conventional PT-symmetric system averages the gain and loss to zero in the effective medium (very small frequency) so that it will be difficult to have an exceptional point (i.e. an eigenvalue degeneracy of the constitutive matrix) unless we go to very high values of $\gamma$. In contrast, for our PT-symmetric system where we match $\epsilon$ to $\mu$, the unidirectional reflection can occur at small frequency and small $\gamma$. This is important for experimental realization as easily achievable gain/loss values (small values) can be used by simply working at a longer wavelength. This condition of unidirectional reflection (or an exceptional "curve") is represented by the solid black line approaching the origin. This behavior can be explained analytically by considering the effective medium of such a system (see Appendix for details):

$$C_{\text{eff}}^{-1} = \begin{pmatrix} \dfrac{1+\epsilon_1}{2} & \dfrac{ik_0 a(\epsilon_1 \mu_2 - 1)}{8} \\ -\dfrac{ik_0 a(\epsilon_1 \mu_2 - 1)}{8} & \dfrac{1+\mu_2}{2} \end{pmatrix}, \tag{15}$$

which satisfies the PT-symmetry condition for the effective medium (Eq. (9)) automatically (with $\epsilon_1 = \mu_2^*$). Then, the exceptional point (eigenvalue degeneracy of $C^{-1}$ where $c_1^{-1} = c_2^{-1}$) occurs at the black solid line (in Fig. 3(b)) where

$$k_0 a = \frac{4\gamma}{Re(\epsilon_1)^2 - 1 + \gamma^2}. \tag{16}$$

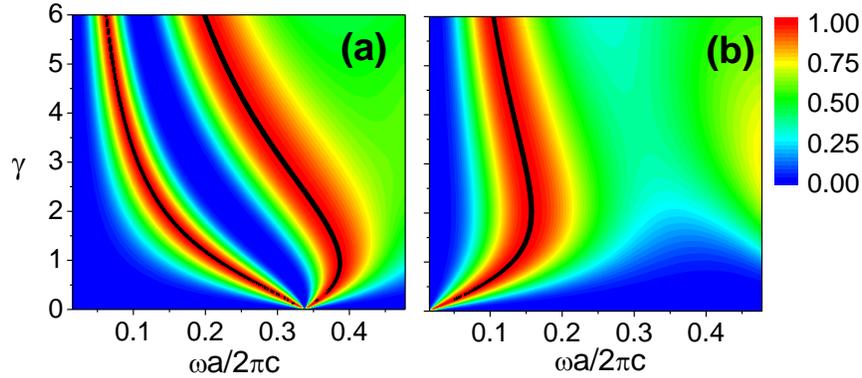

Fig. 3. (Color online) The degree of unidirectionality $\chi$ for PT-symmetric photonic crystals by stacking alternating slabs with permittivity (permeability) $\epsilon_1$ ($\mu_1$) and $\epsilon_2$ ($\mu_2$). (a) Matching alternating slabs with $\epsilon_1 = 2.25 + i\gamma$, $\epsilon_2 = 2.25 - i\gamma$ and $\mu_1 = \mu_2 = 1$. (b) Matching alternating slabs with $\epsilon_1 = 2.25 + i\gamma$, $\mu_2 = 2.25 - i\gamma$ and $\epsilon_2 = \mu_1 = 1$. The black solid lines show the exceptional point (or curve) with unidirectional character in reflection.

Next, we consider the photonic crystal (in Fig. 3(b)) at a fixed freespace wavelength of $\lambda = 4.8$ (with lattice constant taken as $a = 0.08$) in the long wave length limit. Figure 4 (a) shows the eigenvalue behavior of $C^{-1}$ against $\gamma$. The solid lines represent the results from the analytic formula (Eq. (15)), which agree well with the result obtained from Eq. (6) (while $S$ is numerically obtained from transfer matrix method). As we increase $\gamma$ from 0 (the lossless case), at a certain threshold value $\gamma_{ex} = 0.107$, the two eigenvalues coincide at an exceptional point and split into a complex conjugate pair. Figure 4(b) shows the parametric plot of the real and imaginary parts of the eigenvalues of $C^{-1}$ as we increase $\gamma$. The eigenvalues approach and meet at the exceptional point and then rapidly change their directions. The $\gamma$ required for an exceptional point is tunable in this system, as can be seen from Eq. (16). It can be lowered by either using lower permittivity/ permeability slabs (<2.25) or by lowering either $k_0$ or $a$ (i.e. longer wavelength or thinner slab). This can be useful to allow PT-symmetric behavior to be observed experimentally with achievable gain/loss levels. As an example, using a permittivity/permeability with a real part of 1.3 requires a $\gamma$ at around 0.02 for an exceptional point to occur.

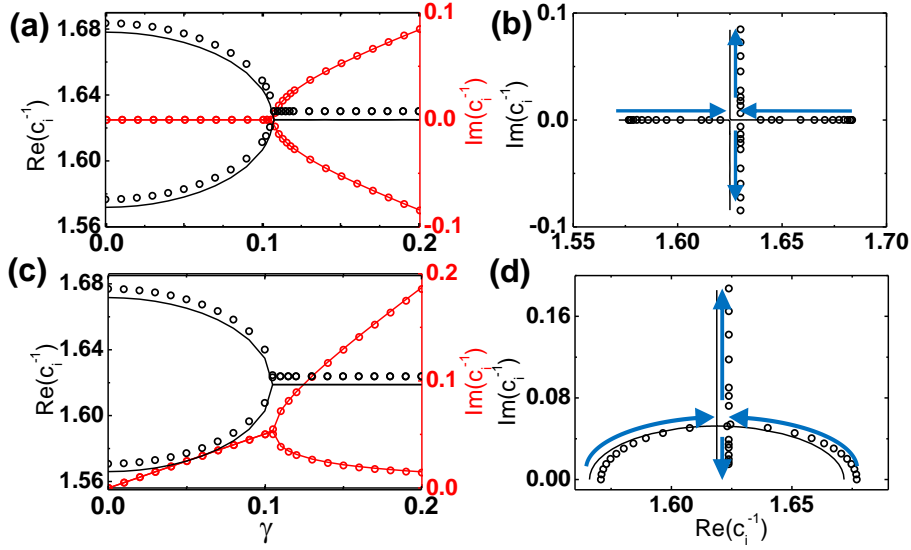

Fig. 4. (Color online) Splitting of eigenvalues ($c_i^{-1}$, $i = 1,2$) of the effective constitutive matrix $C^{-1}$. The scattered symbols (solid lines) are the numerical results from Eq. (6) (analytic effective medium results from Eq. (15)). a) and c) show the real (black) and imaginary (red) parts of the eigenvalues vs. $\gamma$ for the ideal ($\epsilon_1 = \mu_2^*$) and lossy cases (loss only in $\mu_2$) respectively. b) and d) are the corresponding plots of the eigenvalues in the complex plane. The arrows show the direction of increasing $\gamma$.

In the above, we have matched the gain and loss ($\gamma$) to have exactly the same magnitude. To make an experimental realization easier, we can also consider a passive system as a non-ideal PT-symmetric system [10]. If we employ a lossless permittivity of the first slab ($Im(\epsilon_1) = 0$) and only employ loss in the second magnetic slab ($Im(\mu_2) = 2\gamma > 0$), we similarly look for the condition on how we should match the real parts of $\epsilon_1$ and $\mu_2$ so that we can encounter an exceptional point when we vary $\gamma$ upon such a condition. In this case, the condition of the exceptional point existence (by considering eigenvalues of $C^{-1}$ in Eq. (15)) becomes

$$Re(\mu_2) = Re(\epsilon_1)\frac{4 + k_0^2 a^2}{4 + Re(\epsilon_1)^2 k_0^2 a^2}, \tag{17}$$

which is approximately the original working condition for ideal PT-symmetry $Re(\epsilon_1) \cong Re(\mu_1)$ at a small working frequency. In our case, it corresponds to $\epsilon_1 = 2.25$ and $\mu_2 = 2.22528 + 2i\gamma$ (we have used $2i\gamma$ so that the exceptional points occur at roughly the same $\gamma$ [38]). The eigenvalue movements of the non-ideal system are plotted in Fig. 4 c) and d), displaying a PT phase transition again at $\gamma \sim 0.105$, though in this case there is an additional linear loss term due to the absence of gain to cancel out the loss (Fig. 4c).

The results directly obtained from $C^{-1}$ are also in good agreement with our analytic approximation using Eq. (15), though there is a small shift in the real part of the permittivity due to the approximation. As the unidirectional character in reflection is invariant with respect to the number of unit cells and occurs in the long wavelength limit, the same phenomenon can also be equivalently demonstrated by an ultrathin metasurface, which consists of only one unit cell with subwavelength thickness. The passive system consideration also provides an additional flexibility in realizing PT-symmetric metasurfaces by using only passive metamaterial structures.

In this work we have demonstrated a class of PT-symmetric photonic crystals and metamaterials in matching electric permittivity with magnetic permeability. This is achieved by adopting an effective parity operator with the coupling between electric and magnetic fields. We have shown a PT-phase transition of the band structure by studying the eigenvalue problem of the propagation matrix. Our analysis reveals that the reflection from a PT-symmetric photonic crystal has unidirectional character, which is associated with the exceptional point of the constitutive matrix. On the one hand, the proposed PT-symmetry shares some common features with the conventional one in matching only one material potential. On the other hand, the cross-matching of material potentials allows exceptional points to occur in the long wavelength limit so that PT-symmetry metamaterials and metasurfaces can be constructed easily.

**ACKNOWLEDGEMENT**

J.L. would like to acknowledge funding support from the European Union's Seventh Framework Programme under Grant Agreement No. 630979. S.R. is supported by the Austrian Science Fund (FWF) through Projects No. SFB NextLite F49-10 and No. I 1142- N27 (GePartWave).

**Appendix A: Effective medium with bianisotropy from stacking alternating layers at finite frequency.**

For periodic AB layers in the *z*-direction, suppose we only consider propagation along the *z*-direction with volume filling ratio $f_A$ and $f_B = 1 - f_A$, layer A (B) has isotropic permittivity $\epsilon_A$ ($\epsilon_B$) and permeability $\mu_A$ ($\mu_B$). In the quasi-static limit, the effective medium (along transverse direction which the E-field and H-field of the propagating wave exhibits) is given by

$$\epsilon_{\text{eff}} = f_A \epsilon_A + f_B \epsilon_B, \qquad (A.1)$$

and

$$\mu_{\text{eff}} = f_A\mu_A + f_B\mu_B, \tag{A.2}$$

using simple volume averaging. Then, we can employ the complex transmission and reflection amplitude given by the S-matrix in Eq. (6). It can be applied on the two individual A and B layers and then be compared to the complex transmission and reflection amplitudes of the total double slab. Using the same equation, we can then obtain the effective $C^{-1}$ matrix. The diagonal values (permittivity and permeability) are still governed by Eq. (A.1) and (A.2) while the off-diagonal value in the $C^{-1}$ matrix (bianisotropy here) is represented by

$$\kappa_{\text{eff}} = \frac{1}{2}k_0 a f_A f_B (\epsilon_B \mu_A - \epsilon_A \mu_B), \tag{A.3}$$

where $a$ is the periodicity. The bianisotropy term is small but cannot be neglected here at a finite frequency, e.g. for metamaterials. We note that the above result can be extended in a straight-forward manner if the two layers are now replaced by two anisotropic materials sharing the same set of principal axes along the *x*- and *y*-directions since the TE and TM polarizations are decoupled.